\documentclass[10pt,letter,prl,twocolumn,hypertext]{revtex4-1}
\usepackage{ifthen}
\newboolean{uprightparticles}
\setboolean{uprightparticles}{false} 

\usepackage{amssymb}
\usepackage{amsfonts}
\usepackage{upgreek}
\usepackage{xspace}
\usepackage{color}
\usepackage{amsmath}
\usepackage{graphicx}
\usepackage{bm}

\newboolean{articletitles}
\setboolean{articletitles}{true}

\newboolean{pdflatex}
\setboolean{pdflatex}{true} 

\def\lhcb {LHCb\xspace}
\def\ux85 {UX85\xspace}

\def\babar  {BaBar\xspace}
\def\belle  {Belle\xspace}

\def\cdf    {CDF\xspace}

\ifthenelse{\boolean{uprightparticles}}%
{

 \def\Pmu         {\ensuremath{\upmu}\xspace}

 \def\Ppsi        {\ensuremath{\uppsi}\xspace}

 \def\PDelta      {\ensuremath{\Delta}\xspace}                 
 \def\PXi      {\ensuremath{\Xi}\xspace}                 
 \def\PLambda      {\ensuremath{\Lambda}\xspace}                 
 \def\PSigma      {\ensuremath{\Sigma}\xspace}                 
 \def\POmega      {\ensuremath{\Omega}\xspace}                 
 \def\PUpsilon      {\ensuremath{\Upsilon}\xspace}                 
 
 \def\PB      {\ensuremath{\mathrm{B}}\xspace}                 
                  
 \def\PD      {\ensuremath{\mathrm{D}}\xspace}

 \def\PJ      {\ensuremath{\mathrm{J}}\xspace}                 
 \def\PK      {\ensuremath{\mathrm{K}}\xspace}

 \def\Pb      {\ensuremath{\mathrm{b}}\xspace}

 \def\Pi      {\ensuremath{\mathrm{i}}\xspace}

 \def\Ps      {\ensuremath{\mathrm{s}}\xspace}

}
{

 \def\Pmu         {\ensuremath{\mu}\xspace}

 \def\Ppsi        {\ensuremath{\psi}\xspace}                 
                  
 \mathchardef\PDelta="7101
 \mathchardef\PXi="7104
 \mathchardef\PLambda="7103
 \mathchardef\PSigma="7106
 \mathchardef\POmega="710A
 \mathchardef\PUpsilon="7107
                  
 \def\PB      {\ensuremath{B}\xspace}                 
                  
 \def\PD      {\ensuremath{D}\xspace}

 \def\PJ      {\ensuremath{J}\xspace}                 
 \def\PK      {\ensuremath{K}\xspace}

 \def\Pb      {\ensuremath{b}\xspace}

 \def\Pi      {\ensuremath{i}\xspace}

 \def\Ps      {\ensuremath{s}\xspace}

}

\def\mup        {\ensuremath{\Pmu^+}\xspace}
\def\mun        {\ensuremath{\Pmu^-}\xspace} 
\def\mumu       {\ensuremath{\Pmu^+\Pmu^-}\xspace}

\def\squark    {\ensuremath{\Ps}\xspace}

\def\bquark    {\ensuremath{\Pb}\xspace}

\def\kaon  {\ensuremath{\PK}\xspace}
  \def\Kbar  {\kern 0.2em\overline{\kern -0.2em \PK}{}\xspace}

\def\Kz    {\ensuremath{\kaon^0}\xspace}
\def\Kzb   {\ensuremath{\Kbar^0}\xspace}
\def\KzKzb {\ensuremath{\Kz \kern -0.16em \Kzb}\xspace}
\def\Kp    {\ensuremath{\kaon^+}\xspace}
\def\Km    {\ensuremath{\kaon^-}\xspace}

\def\KpKm  {\ensuremath{\Kp \kern -0.16em \Km}\xspace}

\def\Kstarz  {\ensuremath{\kaon^{*0}}\xspace}

  \def\Dbar    {\kern 0.2em\overline{\kern -0.2em \PD}{}\xspace}
\def\D       {\ensuremath{\PD}\xspace}

\def\Dz      {\ensuremath{\D^0}\xspace}
\def\Dzb     {\ensuremath{\Dbar^0}\xspace}
\def\DzDzb   {\ensuremath{\Dz {\kern -0.16em \Dzb}}\xspace}
\def\Dp      {\ensuremath{\D^+}\xspace}
\def\Dm      {\ensuremath{\D^-}\xspace}

\def\DpDm    {\ensuremath{\Dp {\kern -0.16em \Dm}}\xspace}

\def\Dstarp  {\ensuremath{\D^{*+}}\xspace}

\def\B       {\ensuremath{\PB}\xspace}
  \def\Bbar    {\kern 0.18em\overline{\kern -0.18em \PB}{}\xspace}

\def\Bu      {\ensuremath{\B^+}\xspace}

\def\Bp      {\ensuremath{\Bu}\xspace}

\def\Bd      {\ensuremath{\B^0}\xspace}
\def\Bs      {\ensuremath{\B^0_\squark}\xspace}

\def\Bdb     {\ensuremath{\Bbar^0}\xspace}

\def\jpsi     {\ensuremath{{\PJ\mskip -3mu/\mskip -2mu\Ppsi\mskip 2mu}}\xspace}

  \def\Y#1S{\ensuremath{\PUpsilon{(#1S)}}\xspace}

\def\BF         {{\ensuremath{\cal B}\xspace}}

\newcommand{\decay}[2]{\ensuremath{#1\!\to #2}\xspace}         

\def\to                 {\ensuremath{\rightarrow}\xspace}

\def\qsq       {\ensuremath{q^2}\xspace}

\def\FL       {\ensuremath{F_{\mathrm{L}}}\xspace}
\def\AT#1     {\ensuremath{A_{\mathrm{T}}^{#1}}\xspace}           

\def\ctl       {\ensuremath{\cos{\theta_l}}\xspace}
\def\ctk       {\ensuremath{\cos{\theta_K}}\xspace}

\def\C#1      {\ensuremath{\mathcal{C}_{#1}}\xspace}                       
\def\Cp#1     {\ensuremath{\mathcal{C}_{#1}^{'}}\xspace}                    
\def\Ceff#1   {\ensuremath{\mathcal{C}_{#1}^{\mathrm{(eff)}}}\xspace}        
\def\Cpeff#1  {\ensuremath{\mathcal{C}_{#1}^{'\mathrm{(eff)}}}\xspace}       
\def\Ope#1    {\ensuremath{\mathcal{O}_{#1}}\xspace}                       
\def\Opep#1   {\ensuremath{\mathcal{O}_{#1}^{'}}\xspace}                    

\newcommand{\tev}{\ensuremath{\mathrm{\,Te\kern -0.1em V}}\xspace}
\newcommand{\gev}{\ensuremath{\mathrm{\,Ge\kern -0.1em V}}\xspace}
\newcommand{\mev}{\ensuremath{\mathrm{\,Me\kern -0.1em V}}\xspace}
\newcommand{\kev}{\ensuremath{\mathrm{\,ke\kern -0.1em V}}\xspace}
\newcommand{\ev}{\ensuremath{\mathrm{\,e\kern -0.1em V}}\xspace}
\newcommand{\gevc}{\ensuremath{{\mathrm{\,Ge\kern -0.1em V\!/}c}}\xspace}
\newcommand{\mevc}{\ensuremath{{\mathrm{\,Me\kern -0.1em V\!/}c}}\xspace}
\newcommand{\gevcc}{\ensuremath{{\mathrm{\,Ge\kern -0.1em V\!/}c^2}}\xspace}
\newcommand{\gevgevcccc}{\ensuremath{{\mathrm{\,Ge\kern -0.1em V^2\!/}c^4}}\xspace}
\newcommand{\mevcc}{\ensuremath{{\mathrm{\,Me\kern -0.1em V\!/}c^2}}\xspace}

\def\invfb   {\ensuremath{\mbox{\,fb}^{-1}}\xspace}

\def\gsim{{~\raise.15em\hbox{$>$}\kern-.85em
          \lower.35em\hbox{$\sim$}~}\xspace}
\def\lsim{{~\raise.15em\hbox{$<$}\kern-.85em
          \lower.35em\hbox{$\sim$}~}\xspace}

\def\pt         {\mbox{$p_{\rm T}$}\xspace}

\def\tell1  {TELL1\xspace}
\def\ukl1   {UKL1\xspace}

\def\BdKstarMuMu     {\decay{\Bd}{\Kstarz \mup\mun}}
\def\BpKMuMu     {\decay{\Bp}{\Kp \mup\mun}}
\def\BsPhiMuMu     {\decay{\Bs}{\phi \mup\mun}}
\def\BdJpsiKstar     {\decay{\Bd}{\jpsi\Kstarz}}
\def\BdpsiKstar     {\decay{\Bd}{\psi(2S)\Kstarz}}
\def\afb{\ensuremath{A_{\mathrm{FB}}}\xspace}
\def\qsq{\ensuremath{q^2}\xspace}
\def\fl{\ensuremath{F_{\mathrm{L}}}\xspace}
\def\mkpimumu{\ensuremath{m_{K^+\pi^-\mu^+\mu^-}}\xspace}

\usepackage{mciteplus}
\begin{document}
\title{Differential branching fraction and angular analysis of the
  decay $\boldsymbol{B^{0} \rightarrow K^{*0} \mu^+ \mu^-}$}
\author{
\begin{center}
  \textbf{(LHCb collaboration)}
\end{center}

R.~Aaij$^{23}$, 
C.~Abellan~Beteta$^{35,n}$, 
B.~Adeva$^{36}$, 
M.~Adinolfi$^{42}$, 
C.~Adrover$^{6}$, 
A.~Affolder$^{48}$, 
Z.~Ajaltouni$^{5}$, 
J.~Albrecht$^{37}$, 
F.~Alessio$^{37}$, 
M.~Alexander$^{47}$
% ...

{\footnotesize \it
% ...
$ ^{5}$Clermont Universit\'{e}, Universit\'{e} Blaise Pascal, CNRS/IN2P3, LPC, Clermont-Ferrand, France\\
$ ^{6}$CPPM, Aix-Marseille Universit\'{e}, CNRS/IN2P3, Marseille, France\\
% ...
$ ^{23}$Nikhef National Institute for Subatomic Physics, Amsterdam, The Netherlands\\
% ...
$ ^{35}$Universitat de Barcelona, Barcelona, Spain\\
$ ^{36}$Universidad de Santiago de Compostela, Santiago de Compostela, Spain\\
$ ^{37}$European Organization for Nuclear Research (CERN), Geneva, Switzerland\\
% ...
$ ^{42}$H.H. Wills Physics Laboratory, University of Bristol, Bristol, United Kingdom\\
% ...
$ ^{47}$School of Physics and Astronomy, University of Glasgow, Glasgow, United Kingdom\\
$ ^{48}$Oliver Lodge Laboratory, University of Liverpool, Liverpool, United Kingdom\\
% ...
$ ^{n}$LIFAELS, La Salle, Universitat Ramon Llull, Barcelona, Spain\\
% ...
}
}
\begin{abstract}
\noindent The angular distributions and the partial branching fraction of the decay \BdKstarMuMu are studied using  an integrated luminosity of $0.37\invfb$ of data collected with the LHCb detector. 
The forward-backward asymmetry of the muons, \afb, the fraction of longitudinal polarisation, \fl, and the partial branching fraction, $\mathrm{d}\BF/\mathrm{d}q^{2}$, are determined as a function of the dimuon invariant mass. 
The measurements are in good agreement with the Standard Model predictions and are the most precise to date.  
In the dimuon invariant mass squared range $1.00-6.00\gev^2/c^4$, the results are $\afb=-0.06\,^{+0.13}_{-0.14} \pm 0.04$, $\fl=0.55\pm 0.10\pm 0.03$ and  $\mathrm{d}\BF/\mathrm{d}q^{2}=(0.42 \pm 0.06\pm 0.03) \times 10^{-7}c^4/\gev^2$. 
In each case, the first error is statistical and the second systematic. 

\begin{center}
\emph{Published in Physical Review Letters 108, 181806 (2012)}
\end{center}

\end{abstract}


\vspace*{-1cm}
\hspace{-9cm}
\mbox{\Large EUROPEAN ORGANIZATION FOR NUCLEAR RESEARCH (CERN)}

\vspace*{0.1cm}
\hspace*{-9cm}
\begin{tabular*}{16cm}{lc@{\extracolsep{\fill}}r}
\ifthenelse{\boolean{pdflatex}}
{\vspace*{-3.2cm}\mbox{\!\!\!\includegraphics[width=.14\textwidth]{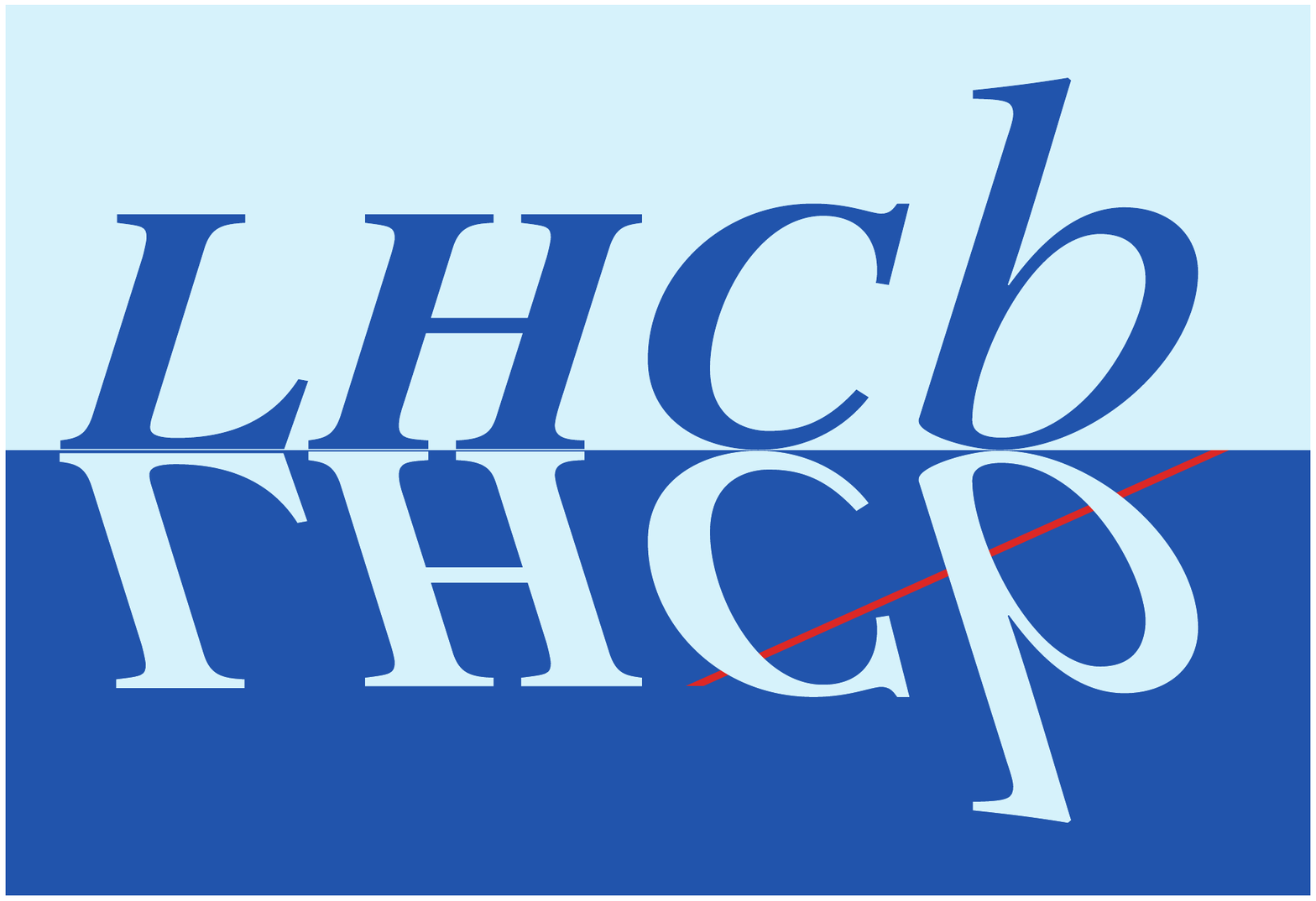}} & &}%
{\vspace*{-1.2cm}\mbox{\!\!\!\includegraphics[width=.12\textwidth]{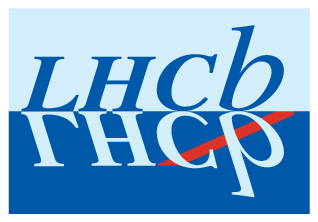}} & &}%
\\
 & & LHCb-PAPER-2011-020 \\
 & & CERN-PH-EP-2011-211 \\ 
 & & \today \\ 
\end{tabular*}
\vspace*{1cm}

\maketitle

\noindent The process \BdKstarMuMu is a flavour changing neutral current decay. In the Standard Model (SM) such decays are suppressed, as they can only proceed via loop processes involving electroweak penguin or box diagrams. 
As-yet undiscovered particles could give additional contributions with comparable amplitudes, and the decay is therefore a sensitive probe of new phenomena.
A number of angular observables in \BdKstarMuMu decays can be theoretically predicted with good control of the relevant form factor uncertainties. 
These include the forward-backward asymmetry of the muons, \afb, and the fraction of longitudinal polarisation, \fl, as functions of the dimuon invariant mass squared, \qsq~\cite{Kruger:1999xa}. 
These observables have previously been measured by the \babar, \belle, and \cdf experiments~\cite{Aubert:2008ju,*PhysRevLett.103.171801,*Aaltonen:2011ja}. 
A more precise determination of \afb is of particular interest as, in the $1.00<\qsq<6.00\gev^2/c^4$ region, previous measurements favour an asymmetry with the opposite sign to that expected in the SM. 
If confirmed, this would be an unequivocal sign of phenomena not described by the SM. 
This letter presents the most precise measurements of \afb, \fl and the partial branching fraction, $\mathrm{d}\BF/\mathrm{d}q^{2}$, to date. The data used for this analysis were taken with the LHCb detector at CERN during 2011 and correspond to an integrated luminosity of $0.37\invfb$. 
The $\Kstarz$ is reconstructed through its decay into the $K^+\pi^-$ final state. 

The \lhcb detector~\cite{Alves:2008zz} is a single-arm spectrometer designed
to study $b$-hadron decays. 
A silicon strip vertex
detector positioned around the interaction region is used to measure the trajectory of charged particles and allows the reconstruction of the primary proton-proton interactions and the displaced secondary vertices characteristic of \B-meson decays. 
A dipole magnetic field and further charged particle tracking stations allow momenta in the range $5<p<100\gevc$ to be determined
 with a precision of $\delta p/p =0.4$--$0.6\%$. 
The experiment has an acceptance for charged particles with pseudorapidity between 2 and 5. 
Two ring imaging Cherenkov (RICH)
detectors allow kaons to be separated from pions or muons over a
momentum range $2<p<100\gevc$. 
Muons 
are
identified on the basis of the number of hits in detectors
interleaved with an iron muon filter. 

The \BdKstarMuMu angular distribution is governed by six \qsq-dependent transversity amplitudes. The decay can be described by \qsq and the three angles $\theta_l,~\theta_K,~\phi$. For the \Bd(\Bdb), $\theta_{l}$ is the angle between the \mup(\mun) and the opposite of the \Bd(\Bdb) direction in the dimuon rest frame, $\theta_{K}$ the angle between the kaon and the direction opposite to the $B$ meson in the \Kstarz rest frame, and $\phi$ the angle between the $\mup\mun$ and $K^+\pi^-$ decay planes in the $B$ rest frame. The inclusion of charge conjugate modes is implied throughout this letter.
At a given \qsq, neglecting the muon mass, the normalised partial differential width integrated over $\theta_K$ and $\phi$ is

\begin{eqnarray}\nonumber
\hspace*{-7mm} \frac{1}{\Gamma} \frac{\mathrm{d}^{2}\Gamma}{\mathrm{d}\cos\theta_{l}\,\mathrm{d}q^{2}} &=& \frac{3}{4}{\fl}(1-\cos^{2}\theta_{l}) +  \\
&& \frac{3}{8}(1-{\fl}) (1+\cos^{2}\theta_{l}) + {\afb}\cos\theta_{l} 
\label{equation:angular:costhetal}
\end{eqnarray}

\noindent and integrated over $\theta_l$ and $\phi$ it is

\begin{eqnarray}\nonumber
\hspace*{-29mm} \frac{1}{\Gamma} \frac{\mathrm{d}^{2}\Gamma}{\mathrm{d}\cos\theta_{K}\,\mathrm{d}q^{2}} &=& \frac{3}{2}{\fl}\cos^{2}\theta_{K} + \\
&& \frac{3}{4}(1-{\fl}) (1-\cos^{2}\theta_{K}).
\label{equation:angular:costhetak}
\end{eqnarray}

\noindent These expressions do not include any broad S-wave contribution to the $\Bd\to K^+\pi^-\mup\mun$ decay and any contribution from low mass tails of higher \Kstarz resonances. 
These contributions are assumed to be small and are neglected in the rest of the analysis.

Signal candidates are isolated from the background using a set of selection criteria which are detailed below. An event-by-event weight is then used to correct for the bias induced by the reconstruction, trigger and selection criteria. 
In order to extract \afb and \fl, simultaneous fits are made to the $K^+\pi^-\mup\mun$ invariant mass distribution and the angular distributions.
The partial branching fraction is measured by comparing the efficiency corrected yield of \BdKstarMuMu decays to the yield of \BdJpsiKstar, where $\jpsi\rightarrow\mup\mun$.

Candidate  \BdKstarMuMu events are first required to pass a hardware trigger which selects muons with a  transverse momentum, $\pt>1.48\gevc$. 
In the subsequent software trigger, at least one of the final state particles is required to have 
both $\pt>0.8\gevc$ and impact parameter $>100~\mu$m with respect to all of the primary proton-proton interaction vertices in the event~\cite{Gligorov:1300771}.
Finally, the tracks of two or more of the final state particles are required to form a vertex which is significantly displaced from the primary vertices in the event~\cite{hlt2toponote}.

In the final event selection, candidates with $K^+\pi^-\mup\mun$ invariant mass in the range \mbox{$5100 < \mkpimumu <5600\mevcc$} and $K^+\pi^-$ invariant mass in the range \mbox{$792<m_{K^+\pi^-}<992\mevcc$} are accepted.
Two types of backgrounds are then considered: combinatorial backgrounds, where the particles selected do not come from a single $b$-hadron decay; and peaking backgrounds, where a single decay is selected but with some of the particle types mis-identified. In addition, the decays \mbox{\BdJpsiKstar} and \mbox{\BdpsiKstar}, where \mbox{$\jpsi,\psi(2S)\rightarrow\mup\mun$}, are removed by rejecting events with dimuon invariant mass, $m_{\mup\mun}$, in the range \mbox{$2946<m_{\mup\mun}<3176\mev/c^2$} or \mbox{$3586<m_{\mup\mun}<3776\mev/c^2$}. 

The combinatorial background, which is smoothly distributed in the reconstructed $K^+\pi^-\mup\mun$ invariant mass, is reduced using a Boosted Decision Tree (BDT). The BDT uses information about the event kinematics, vertex and track quality, impact parameter and particle identification information from the RICH and muon detectors.
The variables that are used in the BDT are chosen so as to induce the minimum possible distortion in the angular and \qsq distributions. For example, no additional requirement is made on the \pt of both of the muons as, at low \qsq, this would remove a large proportion of events with $|\ctl|\sim1$. 
The BDT is trained entirely on data, using samples that are independent of that which is used to make the measurements: triggered and fully reconstructed \BdJpsiKstar events are used as a proxy for the signal decay, and events from the upper \BdKstarMuMu mass sideband ($5350 < m_{K^+\pi^-\mup\mun} < 5600\mevcc$) are used as a background sample. 
The lower mass sideband is not used, as it contains background events formed from partially reconstructed $B$ decays. These events make a negligible contribution in the signal region and have properties different from the combinatorial background which is the dominant background in this region. 

A cut is made on the BDT output in order to optimise the sensitivity to \afb averaged over all \qsq. 
The selected sample has a signal-to-background ratio of three to one.

Peaking backgrounds from \BsPhiMuMu (where \mbox{$\phi\to K^+K^-$}), \BdJpsiKstar and \BdpsiKstar are considered and reduced with a set of vetoes. 
In each case, for the decay to be a potential signal candidate, at least one particle needs to be misidentified. 
For example, \BdJpsiKstar events where a kaon or pion is swapped for one of the muons, peak around the nominal \Bd mass and evade the \jpsi veto described above.
Vetoes for each of these backgrounds are formed by changing the relevant particle mass hypotheses and recomputing the invariant masses, and by making use of the particle identification information. 
In order to avoid having a strongly peaking contribution to the \ctk angular distribution in the upper mass sideband, \BpKMuMu candidates are removed. Events with $K^+\mup\mun$ invariant mass within $60\mevcc$ of the nominal \Bp mass are rejected.
The vetoes for all of these peaking backgrounds remove a negligible amount of signal. 

\begin{figure}[t]
\centering
\ifthenelse{\boolean{pdflatex}}
{\includegraphics[width=\columnwidth]{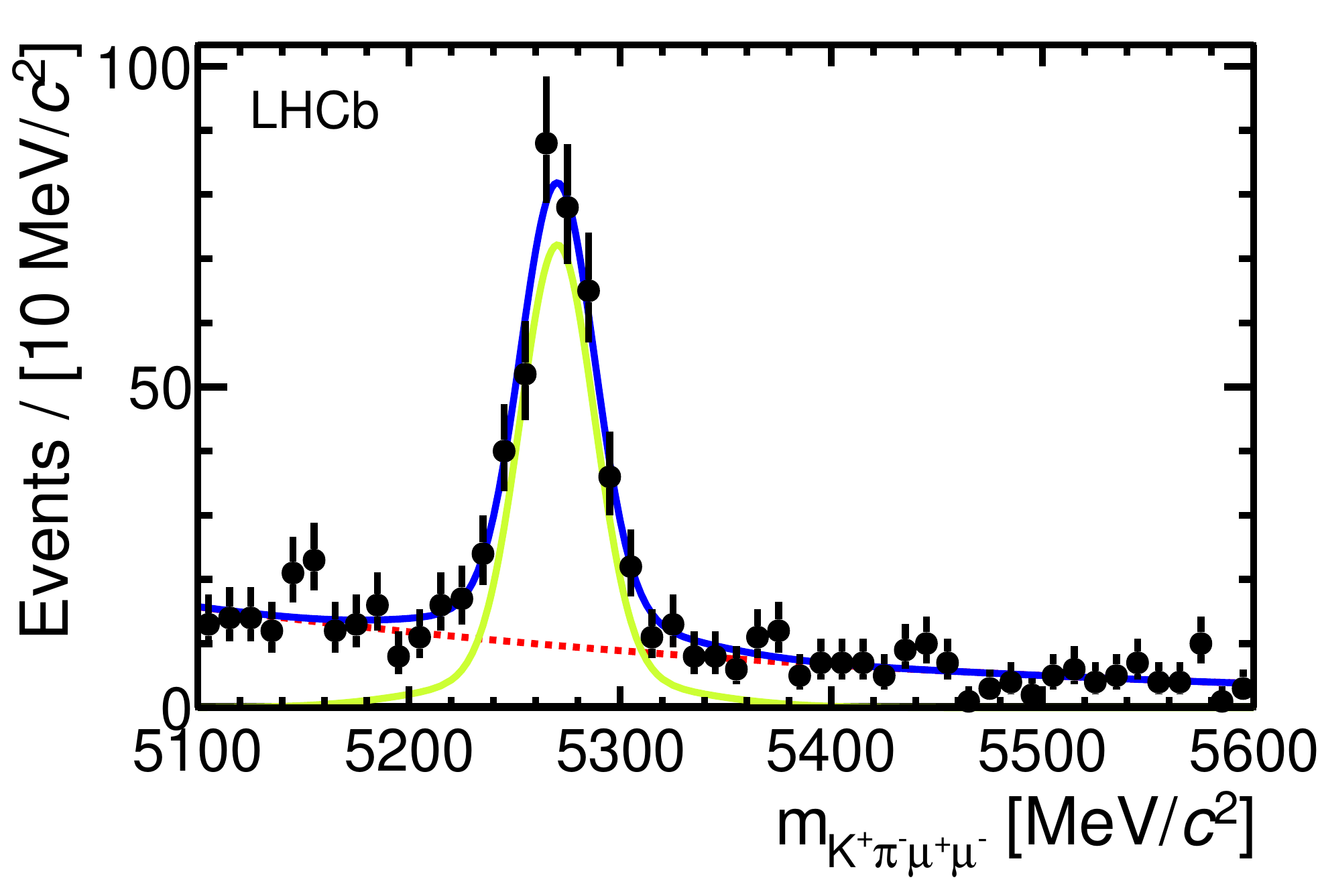}}
{\includegraphics[width=\columnwidth]{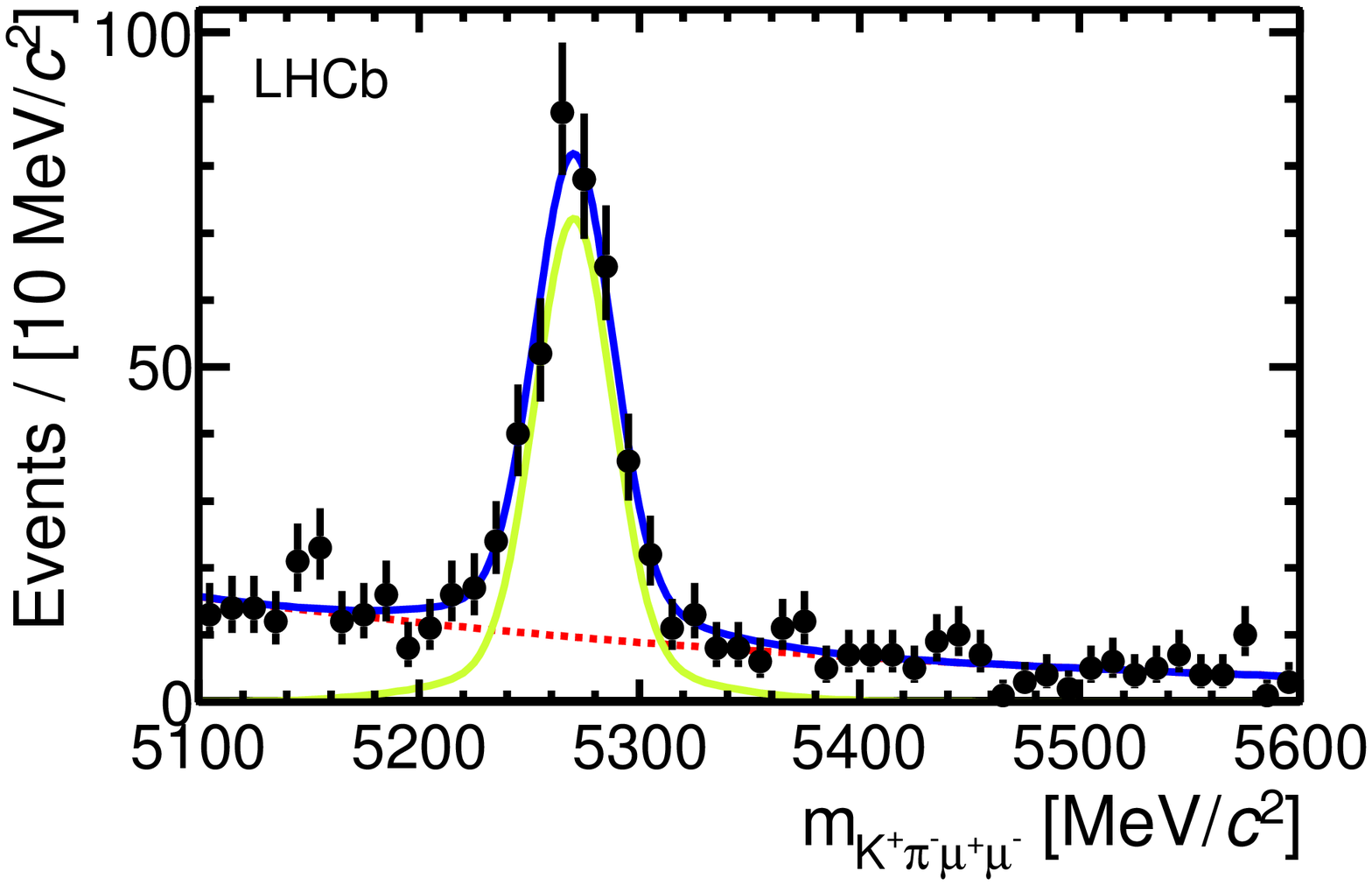}}
\caption{$K^+\pi^-\mup\mun$ invariant mass distribution after the
  application of the full selection as data points with the fit overlaid. The signal component is the green (light) line, the background the red (dashed) line and the full distribution the blue (dark) line.
\label{fig:eventsel:massplot}
}
\end{figure}

After the application of the BDT cut and the above vetoes, a fit is made to 
the $K^+\pi^-\mup\mun$ invariant mass distribution in the entire accepted mass range (see Fig.~\ref{fig:eventsel:massplot}). A double-Gaussian distribution is used for the signal mass shape and an exponential function for the background.
The signal shape is fixed from data using a fit to the \BdJpsiKstar mass peak. 
In the full \qsq range, in a signal mass window of $\pm 50\mevcc$ ($\pm 2.5\sigma$) around the measured \Bd mass, the fit gives an estimate of 
 $337\pm21$ signal events with a background of $97\pm6$ events.
 
The residual peaking background is estimated using simulated events. As detailed below, the accuracy of the simulation is verified by comparing the particle (mis-) identification probabilities with those derived from control channels selected from the data. 
The residual peaking backgrounds are reduced to a level of $6.1$ events, i.e. $1.8\%$
 of the 337 observed signal events. The backgrounds from \BsPhiMuMu and \BdJpsiKstar decays do not give rise to any forward-backward asymmetry and are ignored. 
However, in addition to the above backgrounds, \BdKstarMuMu decays with the kaon and pion swapped give rise to a 0.7\% contribution. 
The change in the sign of the particle which is taken to be the kaon results in a \Bd (\Bdb) being reconstructed as a \Bdb (\Bd), therefore changing the sign of \afb for the candidate. 
This misidentification is accounted for in the fit for the angular observables.

The selected \BdKstarMuMu candidates are weighted in order to correct for the effects of the reconstruction, trigger and selection.
The weights are derived from simulated \BdKstarMuMu events and are normalised such that the average weight is one. In order to be independent of the physics model used in the simulation, the weights are computed based on  \ctk, \ctl and \qsq on an event-by-event basis. 
The variation of detector efficiency with the $\phi$ angle is small and ignoring this variation does not bias the measurements.
Only events with $0.10<\qsq<19.00\gev^2/c^4$ are analysed.

Owing to the relatively unbiased selection, 89\% of events have weights between 0.7 and 1.3, and only 3\% of events have a weight above 2. 
The distortions in the distributions of \ctk, \ctl and \qsq that are induced originate from two main sources. Firstly,
in order to pass through the iron muon filter and give hits in the muon stations, tracks must have at least 3\gevc momentum. At low \qsq this removes events with $|\ctl|\sim 1$. This effect stems from the geometry of the \lhcb detector and is therefore relatively easy to model. Secondly,
events with $\ctk\sim 1$, and hence a slow pion, are removed both by the pion reconstruction and by the impact parameter requirements used in the trigger and BDT selection. 

A number of control samples are used to verify the simulation quality and to correct for differences with respect to the data.
The reproduction of the \Bd momentum and pseudorapidity distributions is verified using \BdJpsiKstar decays. These decays are also used to check that the 
simulation reproduces the measured properties of selected events.
The hadron and muon (mis-)identification probabilities are adjusted using decays 
where the tested particle type can be determined without the use of the particle identification algorithms.
A tag and probe approach with $\jpsi\rightarrow\mup\mun$ decays is used to isolate a clean sample of genuine muons. The decay $\Dstarp\rightarrow\Dz\pi^+$, where $\Dz\rightarrow K^-\pi^+$, is used to give an unambiguous source of kaons and pions. The statistical precision with which it is possible to make the data/simulation comparison gives rise to a systematic uncertainty in the weights which is evaluated below.

The observables \afb and \fl are extracted in bins of \qsq. In each bin, a simultaneous fit to the $K^+\pi^-\mup\mun$ invariant mass distribution and the \ctk and \ctl distributions is performed.
The angular distributions are fitted in both the signal mass window and in the upper mass sideband which determines the background parameters. 
The angular distributions for the signal are given by Eqs.~\ref{equation:angular:costhetal} and~\ref{equation:angular:costhetak} and a second order polynomial in \ctk and in \ctl is used for the background. 

In order to obtain a positive probability density function 
over the entire angular range, Eqs.~\ref{equation:angular:costhetal} and~\ref{equation:angular:costhetak} imply that the conditions $\left| \afb \right| \leq \frac{3}{4}(1-\fl)$ and $0<\fl<1$ must be satisfied. 
To account for this, the maximum likelihood values for \afb and \fl are extracted by performing a profile-likelihood scan over the allowed range. 
The uncertainty on the central value of \afb and \fl is calculated by 
integrating the probability density extracted from the likelihood, assuming a flat prior in \afb and \fl, inside the allowed range. This gives an (asymmetric) 68\% confidence interval.

The partial branching fraction is measured in each of the \qsq bins from a fit to the efficiency corrected $K^+\pi^-\mup\mun$ mass spectrum. The efficiencies are determined relative to the \BdJpsiKstar decay which is used as a normalisation mode.

The event weighting and fitting procedure is validated by fitting the angular distribution of \mbox{\BdJpsiKstar} events, where the physics parameters are known from previous measurements~\cite{Aubert:2007hz}. The product of the \mbox{\BdJpsiKstar} and \mbox{$\jpsi\rightarrow\mup\mun$}
branching fractions is \mbox{$\sim75$} times larger than the branching fraction of \mbox{\BdKstarMuMu}, allowing a precise test of the procedure to be made. 
Fitting the \mbox{\BdJpsiKstar} angular distribution, weighted according to the event-by-event procedure described above, yields values for \fl and \afb in good agreement with those found previously.

\begin{figure}
\centering
\ifthenelse{\boolean{pdflatex}}
{\includegraphics[width=\columnwidth]{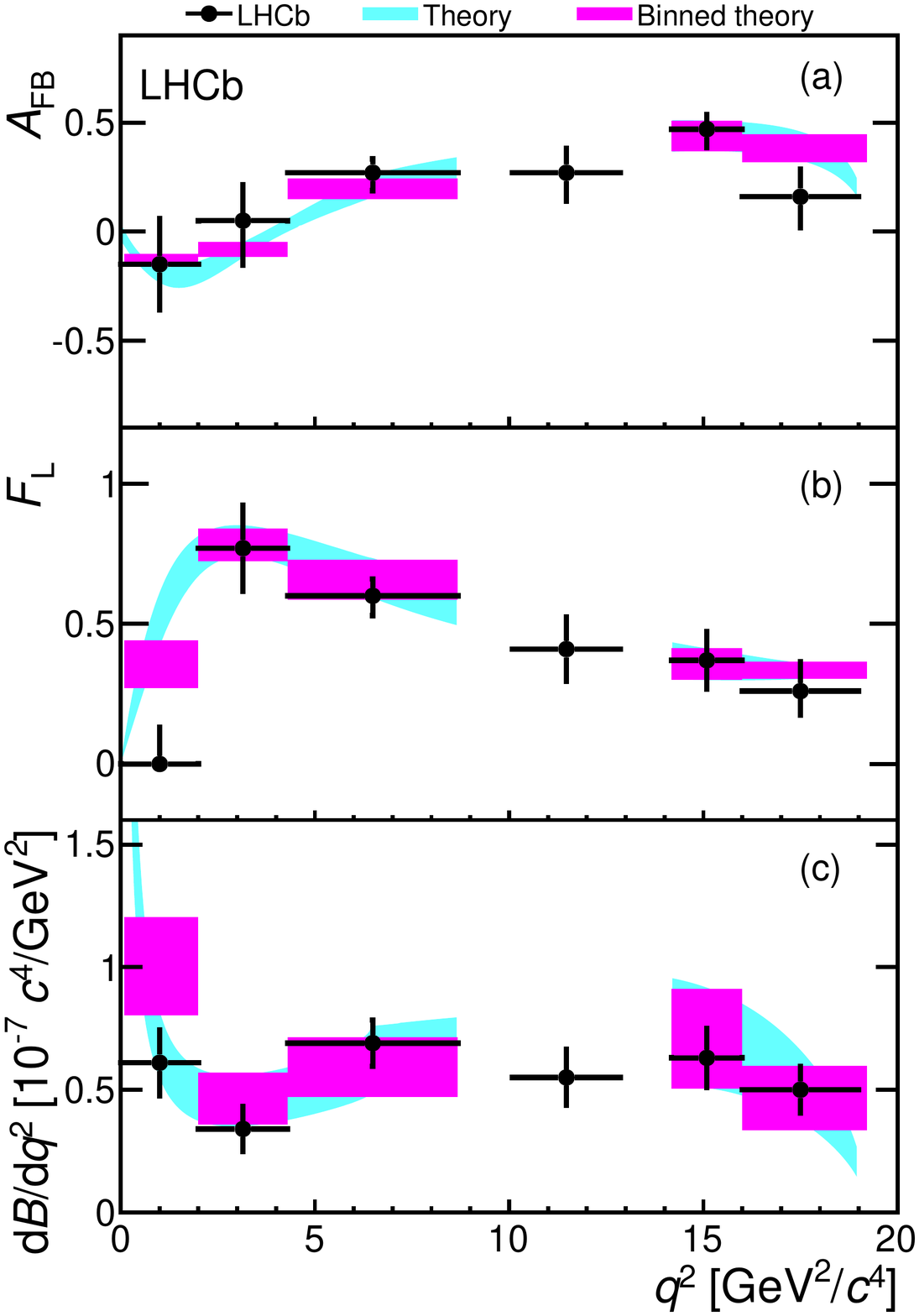}}
{\includegraphics[width=\columnwidth]{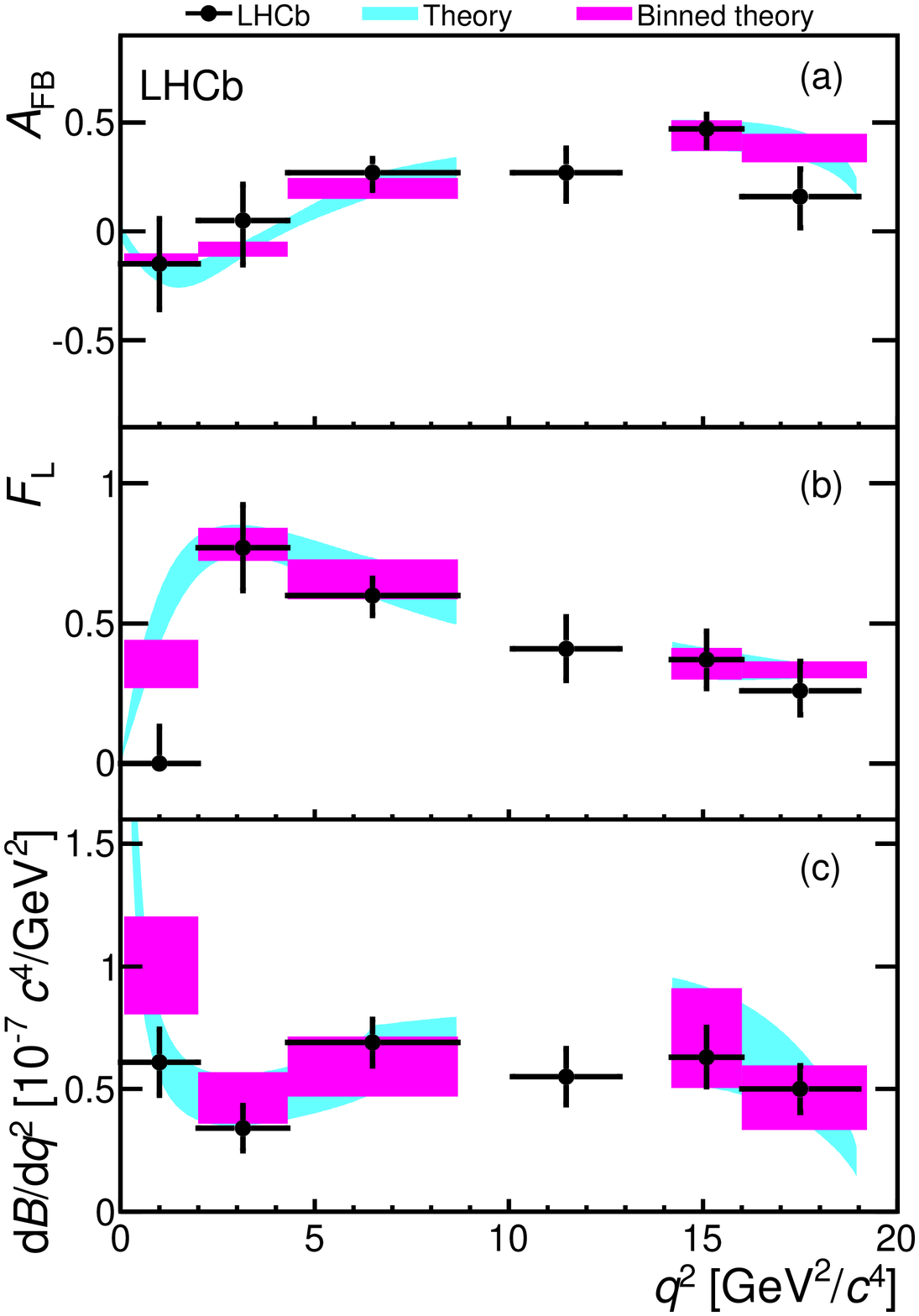}}
\caption{\afb, \fl and $\mathrm{d}\BF/\mathrm{d}q^{2}$ as a function of $q^{2}$. The SM prediction is given by the cyan (light) band, and this prediction rate-averaged across the \qsq\ bins is indicated by the purple (dark) regions. No SM prediction is shown for the region between the two regimes in which the theoretical calculations are made (see text). 
\label{fig:results:lhcbonly}
}
\end{figure}

For \BdKstarMuMu, the fit results for \afb, \fl and
$\mathrm{d}\BF/\mathrm{d}q^{2}$ are shown in
Fig.~\ref{fig:results:lhcbonly} and are tabulated together with the
signal and background yields in Table~\ref{table:result:statplussyst}.
The fit projections are given in the appendix.
Signal candidates are observed in each \qsq bin with more than
$5\sigma$ significance. The compatibility of the fits and the data are assessed using a binned $\chi^2$ test and all fits are found to be of good quality. The measurements in all three quantities are more precise than those of previous experiments and are in good agreement with the SM predictions. 
The predictions are taken from Ref.~\cite{Bobeth:2011gi}. In the low \qsq region they rely on the
factorisation approach~\cite{Beneke:2001at}, which loses accuracy when
approaching the \jpsi resonance; in the high \qsq region, an operator
product expansion in the inverse $b$-quark mass, $1/m_\bquark$, and in $1/\sqrt{\qsq}$ is used~\cite{Grinstein:2004vb}, which is only valid
above the open charm threshold. In both regions the form factor calculations are taken from Ref.~\cite{Ball:2004rg} and a dimensional estimate is made on the uncertainty from 
expansion corrections~\cite{Egede:2008uy}.

In the $1.00 < \qsq < 6.00 \gev^{2}/c^4$ region, the fit gives $\afb = -0.06\,^{+0.13}_{-0.14}\pm 0.04$, $\fl = 0.55\pm 0.10\pm 0.03$ and $\mathrm{d}\BF/\mathrm{d}q^{2} = (0.42 \pm 0.06\pm 0.03) \times 10^{-7}c^4/\gev^2$, where the first error is statistical and the second systematic. The theoretical predictions in the same \qsq range are $\afb = -0.04\pm 0.03$, $\fl = 0.74\,^{+0.06}_{-0.07}$ and $\mathrm{d}\BF/\mathrm{d}q^{2} = (0.50\,^{+0.11}_{-0.10}) \times 10^{-7}c^4/\gev^2$.
The LHCb \afb measurement is a factor $1.5-2.0$ more precise than previous
measurements from the Belle, CDF and BaBar
collaborations~\cite{Aubert:2008ju,*PhysRevLett.103.171801,*Aaltonen:2011ja}
which are, respectively, $\afb=0.26^{+0.27}_{-0.30}\pm 0.07$,
$\afb=0.29^{+0.20}_{-0.23}\pm 0.07$ and, for $\qsq<6.25\gev^2/c^4$,
$\afb=0.24^{+0.18}_{-0.23}\pm 0.05$.
The positive value of \afb preferred in the $1.00<\qsq<6.00\gev^2/c^4$
range in these previous measurements is not favoured by the LHCb data.
The previous measurements of \fl in the same \qsq regions are
$\fl=0.67\pm 0.23 \pm 0.05$ (Belle), $\fl=0.69^{+0.19}_{-0.21}\pm 0.08$
(CDF) and $\fl=0.35\pm0.16 \pm 0.04$ (BaBar). These are in good
agreement with the LHCb result. 

\begin{center}
\begin{table*}
\centering
\caption{
Central values with statistical and systematic uncertainties for \afb, \fl and  $\mathrm{d}\BF/\mathrm{d}q^{2}$ as a function of \qsq. 
The \BdKstarMuMu signal and background yields in the $\pm50\mevcc$ signal mass window with their statistical uncertainties are also indicated, together with the statistical significance of the signal peak that is observed. 
\label{table:result:statplussyst} \\ 
}
\begin{tabular}{|c|c|c|c|c|c|c|}
\hline
$\qsq$ & \afb & \fl & $\mathrm{d}\BF/\mathrm{d}q^{2} $ & Signal & Background & Significance \\ 
$( \gev^{2}/c^4 )$ & & & $(\times 10^{-7} ~c^{4}/\gev^{2})$ & yield & yield & ($\sigma$) \\
\hline
$0.10 < q^{2} < 2.00$ & $-0.15\pm 0.20\pm 0.06$ & $0.00~\,^{+\,0.13}_{-\,0.00}\,\pm 0.02$ & $0.61 \pm 0.12\pm 0.06$ & $48.6 \pm 8.1$ & $16.2 \pm 2.3\phantom0$  & \phantom08.6 \\
$2.00 < q^{2} < 4.30$ & $\phantom-0.05~\,^{+\,0.16}_{-\,0.20}\,\pm 0.04$ & $0.77\pm 0.15 \pm 0.03$ & $0.34 \pm 0.09\pm 0.02$ & $26.5 \pm 6.5$ & $15.7 \pm 2.2\phantom0$  & \phantom05.4 \\
$4.30 < q^{2} < 8.68$ & $\phantom-0.27~\,^{+\,0.06}_{-\,0.08}\,\pm 0.02$ & $0.60~\,^{+\,0.06}_{-\,0.07}\,\pm 0.01$ & $0.69 \pm 0.08\pm 0.05$ & $104.7 \pm 11.9$ & $31.7 \pm 3.3\phantom0$ & 12.4 \\
$10.09 < q^{2} < 12.86$ &$\phantom-0.27~\,^{+\,0.11}_{-\,0.13}\,\pm 0.02$ & $0.41\pm 0.11 \pm 0.03$ & $0.55 \pm 0.09\pm 0.07$ & $62.2 \pm 9.2$ & $20.4 \pm 2.6\phantom0$ & \phantom09.6 \\
$14.18 < q^{2} < 16.00$ & $\phantom-0.47~\,^{+\,0.06}_{-\,0.08}\,\pm 0.03$ & $0.37\pm 0.09\pm 0.05$ & $0.63 \pm 0.11\pm 0.05$ & $44.2 \pm 7.0$ & $4.2 \pm 1.3$ & 10.2  \\
$16.00 < q^{2} < 19.00$ & $\phantom-0.16~\,^{+\,0.11}_{-\,0.13}\,\pm 0.06$ & $0.26~\,^{+\,0.10}_{-\,0.08}\,\pm 0.03$ & $0.50 \pm 0.08\pm 0.05$ & $53.4 \pm 8.1$& $7.0 \pm 1.7$ & \phantom09.8 \\
\hline
$1.00 < q^{2} < 6.00$ & $-0.06~\,^{+\,0.13}_{-\,0.14}\,\pm 0.04$ & $0.55\pm 0.10\pm 0.03$ & $0.42 \pm 0.06\pm 0.03$ & $\phantom076.5 \pm 10.6$& $33.1 \pm 3.2\phantom0$ & \phantom09.9 \\
\hline
\end{tabular}
\end{table*}
\end{center}

For the determination of \afb and \fl, 
the dominant systematic uncertainties arise from the event-by-event weights which are extracted from simulated events, and from the model used to describe the angular distribution of the background. 
The uncertainty on the event-by-event weights is evaluated by fluctuating these weights within their statistical uncertainties and repeating the fitting procedure.
The uncertainty from the background model which is used is estimated by changing this model to one which uses binned templates from the upper mass sideband rather than a polynomial parameterisation. 

The dominant systematic errors for the determination of  $\mathrm{d}\BF/\mathrm{d}q^{2}$ arise from the uncertainties on the particle identification and track reconstruction efficiencies.
These efficiencies are extracted from control channels and are limited by the relevant sample sizes. 
The systematic uncertainty is estimated by fluctuating the efficiencies within the relevant uncertainties and repeating the fitting procedure. An additional systematic uncertainty of $\sim4\%$ arises from the uncertainty in the \BdJpsiKstar and $\jpsi\to\mumu$ branching fractions~\cite{Nakamura:2010zzi}.

The total systematic error on each of \afb and \fl ($\mathrm{d}\BF/\mathrm{d}q^{2}$) is typically $\sim 30\%$ (50\%) of the statistical error, and hence adds $\sim4\%$ ($\sim11\%$) to the total uncertainty.

In summary, using $0.37\invfb$ of data taken with the LHCb detector during 2011, \afb, \fl and $\mathrm{d}\BF/\mathrm{d}q^{2}$ have been determined for the decay \mbox{\BdKstarMuMu}.
These are the most precise measurements of these quantities to-date. All three observables show good agreement with the SM predictions.

\section*{Acknowledgements}

\noindent We express our gratitude to our colleagues in the CERN accelerator
departments for the excellent performance of the LHC. We thank the
technical and administrative staff at CERN and at the LHCb institutes,
and acknowledge support from the National Agencies: CAPES, CNPq,
FAPERJ and FINEP (Brazil); CERN; NSFC (China); CNRS/IN2P3 (France);
BMBF, DFG, HGF and MPG (Germany); SFI (Ireland); INFN (Italy); FOM and
NWO (The Netherlands); SCSR (Poland); ANCS (Romania); MinES of Russia and
Rosatom (Russia); MICINN, XuntaGal and GENCAT (Spain); SNSF and SER
(Switzerland); NAS Ukraine (Ukraine); STFC (United Kingdom); NSF
(USA). We also acknowledge the support received from the ERC under FP7
and the Region Auvergne.




\bibliographystyle{LHCb}


\bibliography{kstarmumureferences}

\newpage
\clearpage
\onecolumngrid
 
{\noindent\bf\Large Appendix}
\appendix

The following fit projections are published as EPAPS material.

\begin{figure*}
\centering
\ifthenelse{\boolean{pdflatex}}
{\includegraphics[width=0.9\textwidth]{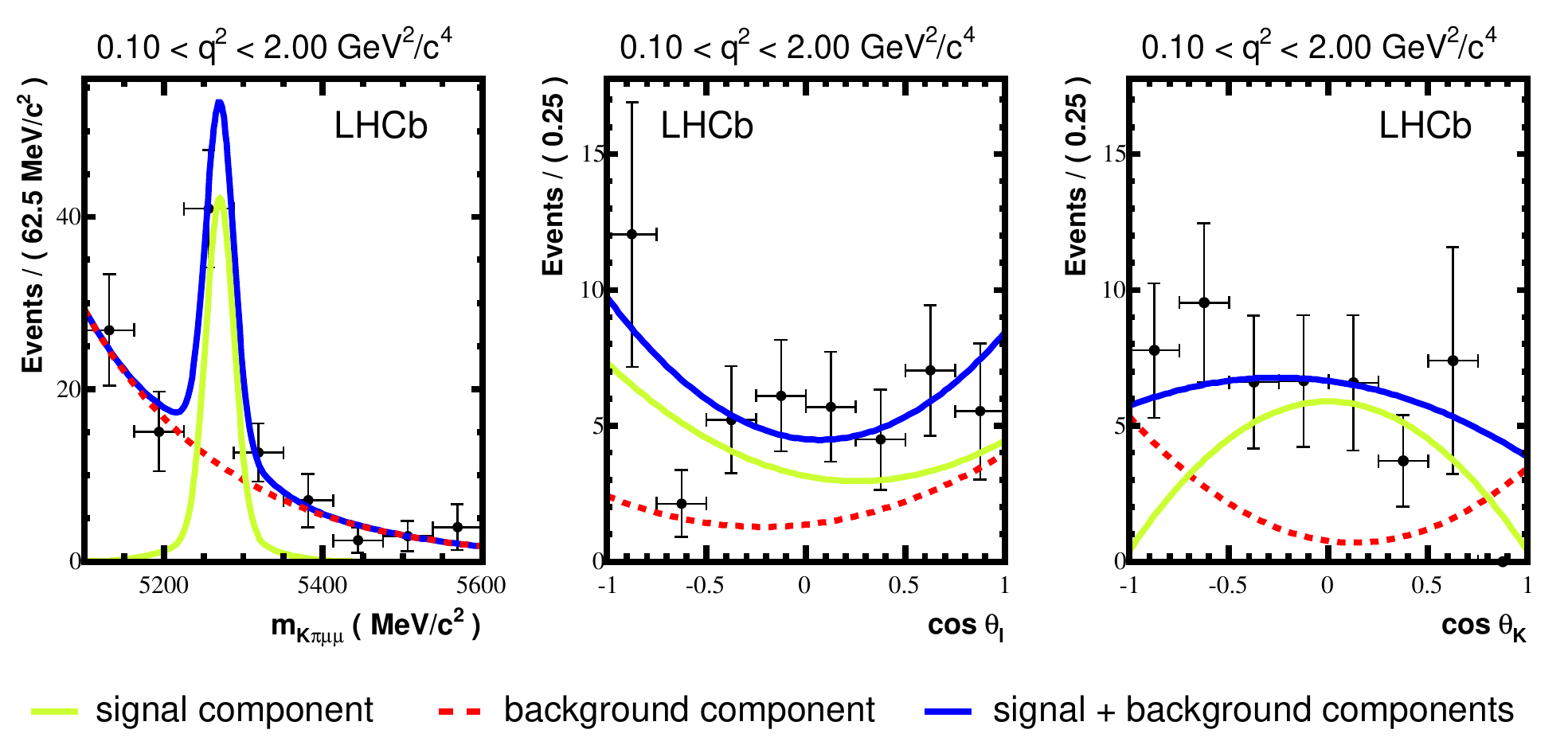}}
{\includegraphics[width=0.9\textwidth]{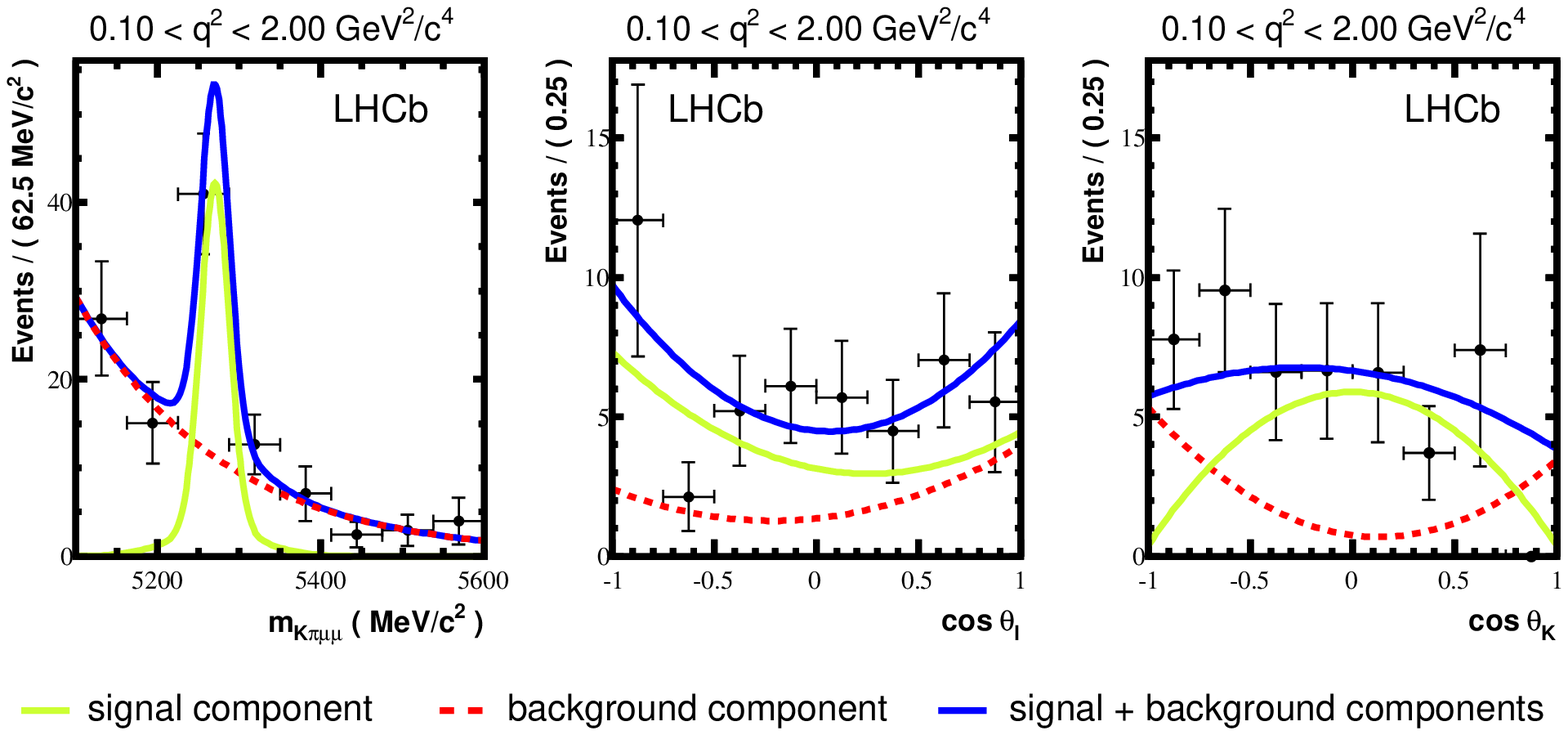}}
\ifthenelse{\boolean{pdflatex}}
{\includegraphics[width=0.9\textwidth]{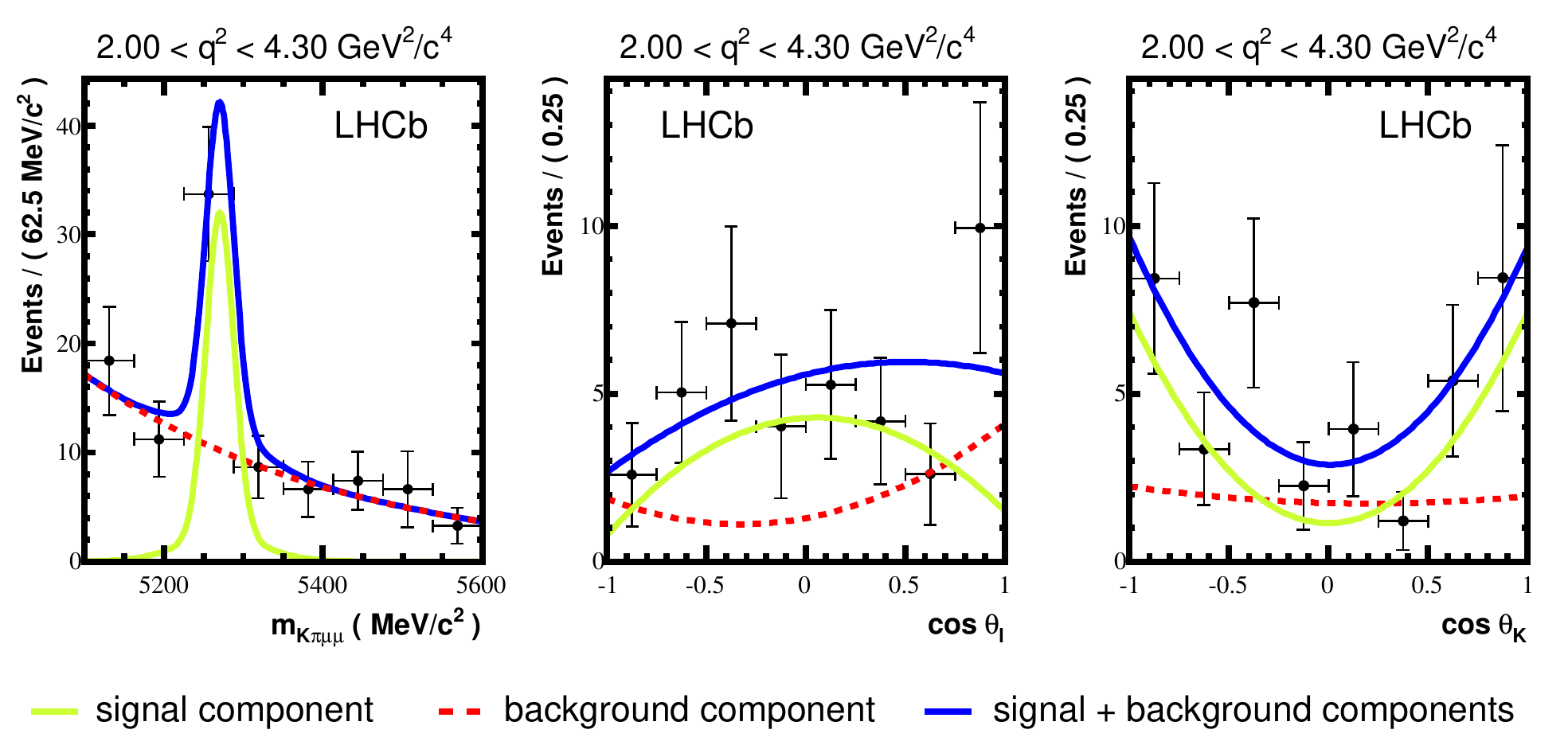}}
{\includegraphics[width=0.9\textwidth]{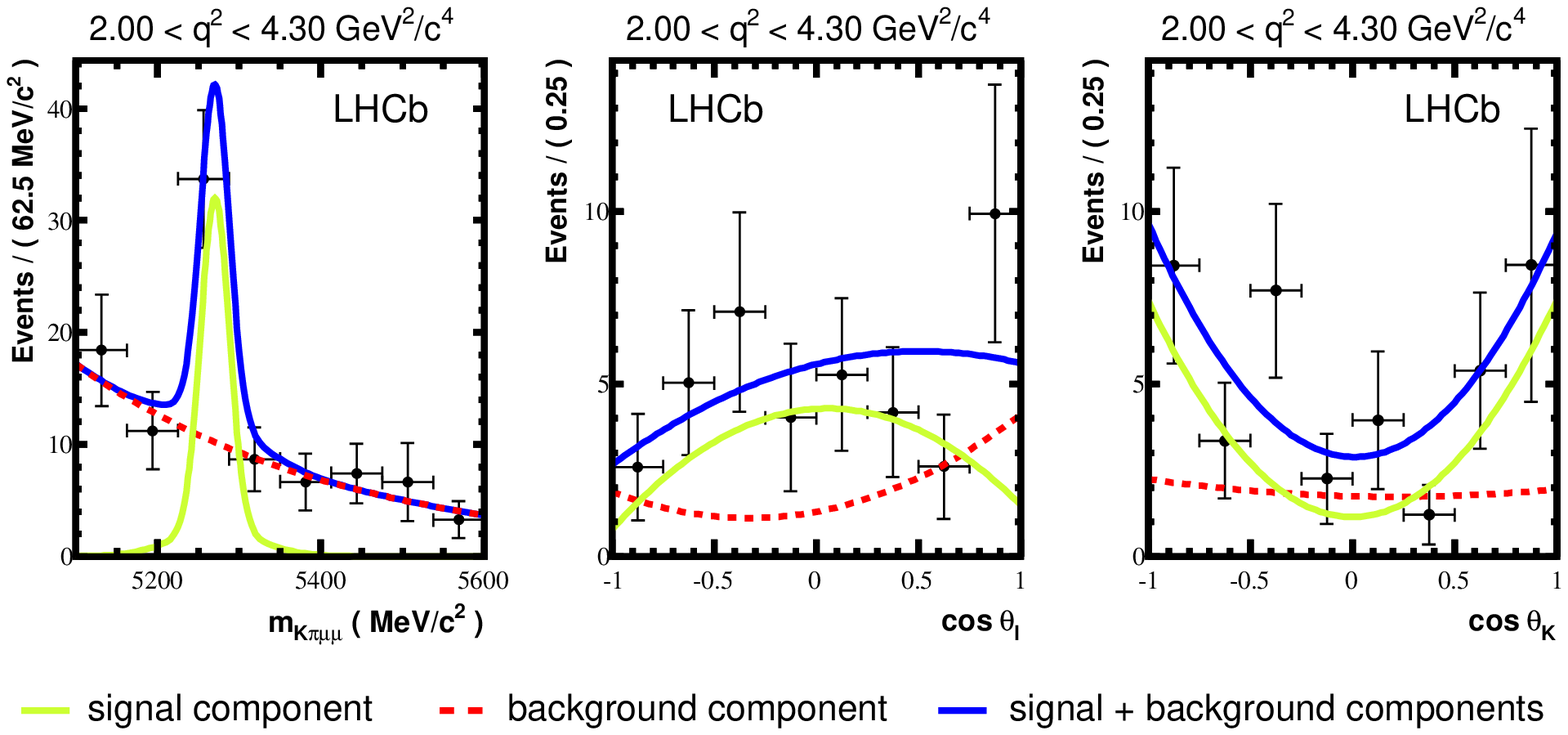}}
\ifthenelse{\boolean{pdflatex}}
{\includegraphics[width=0.9\textwidth]{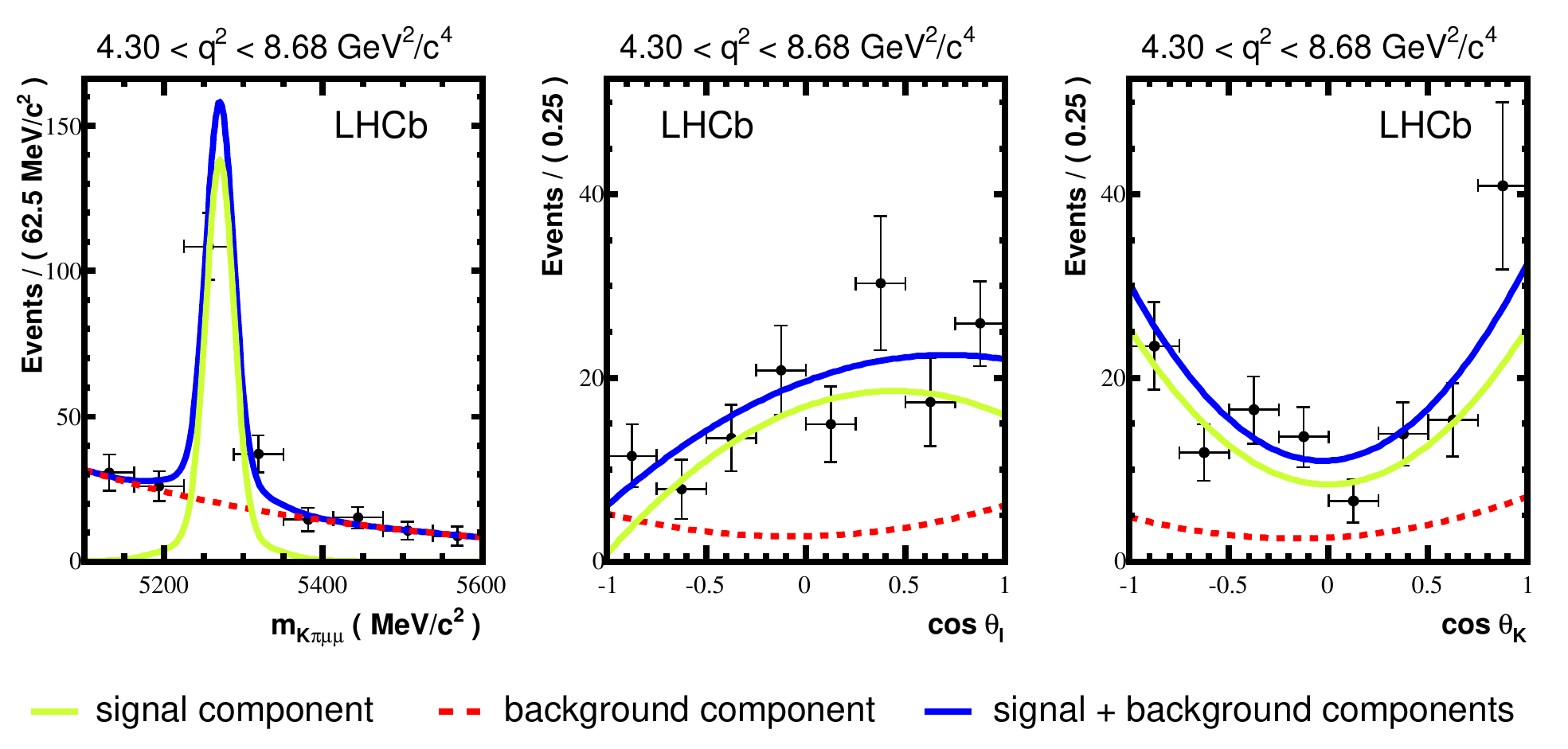}}
{\includegraphics[width=0.9\textwidth]{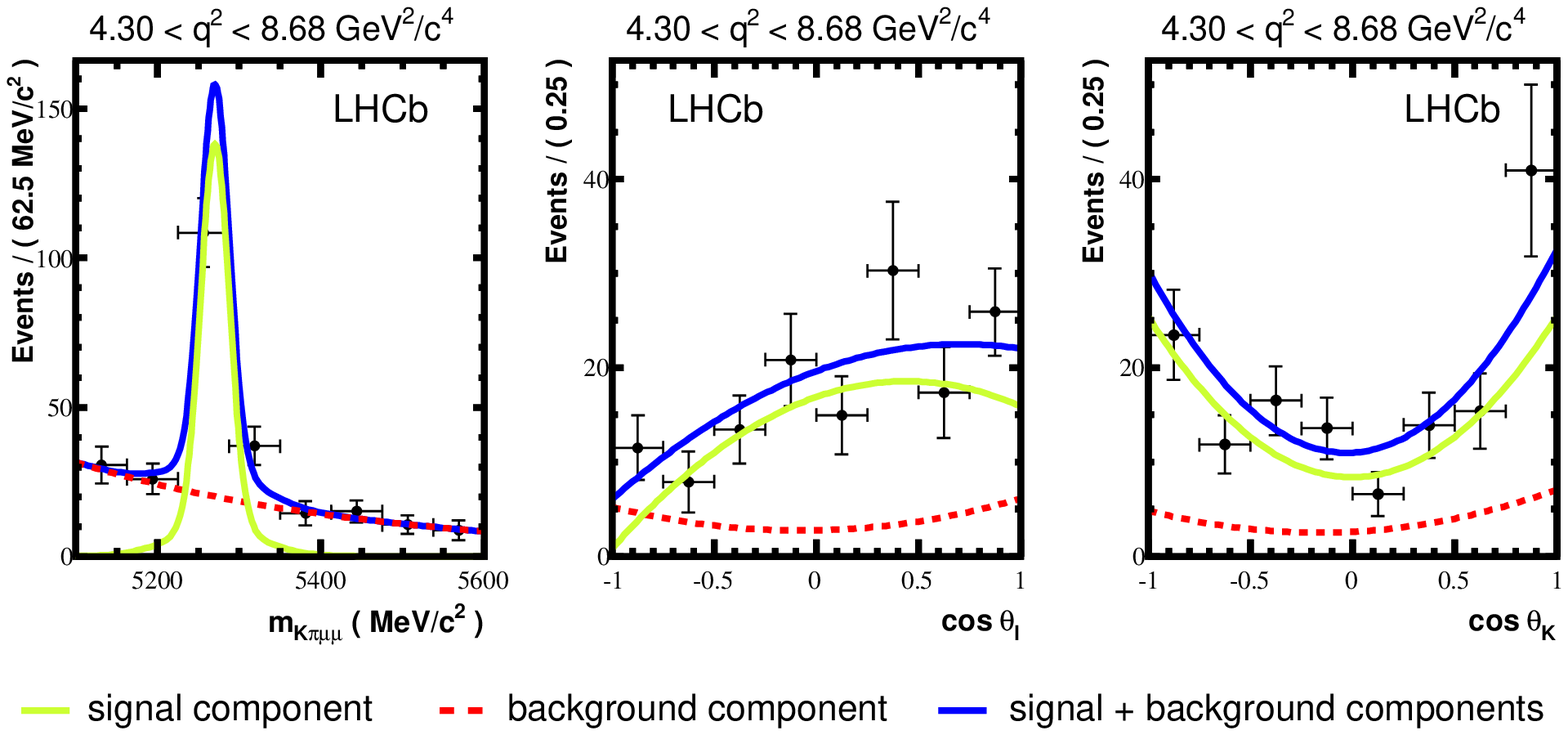}}
\caption{Fit projections for $m_{K\pi\mu\mu}$, \ctl and \ctk for the \qsq bins: $0.10 < q^{2} < 2.00$, $2.00 < q^{2} < 4.30$ and $4.30 < q^{2} < 8.68\gev^{2}/c^4$.}
\end{figure*}

\begin{figure*}
\centering
\ifthenelse{\boolean{pdflatex}}
{\includegraphics[width=0.9\textwidth]{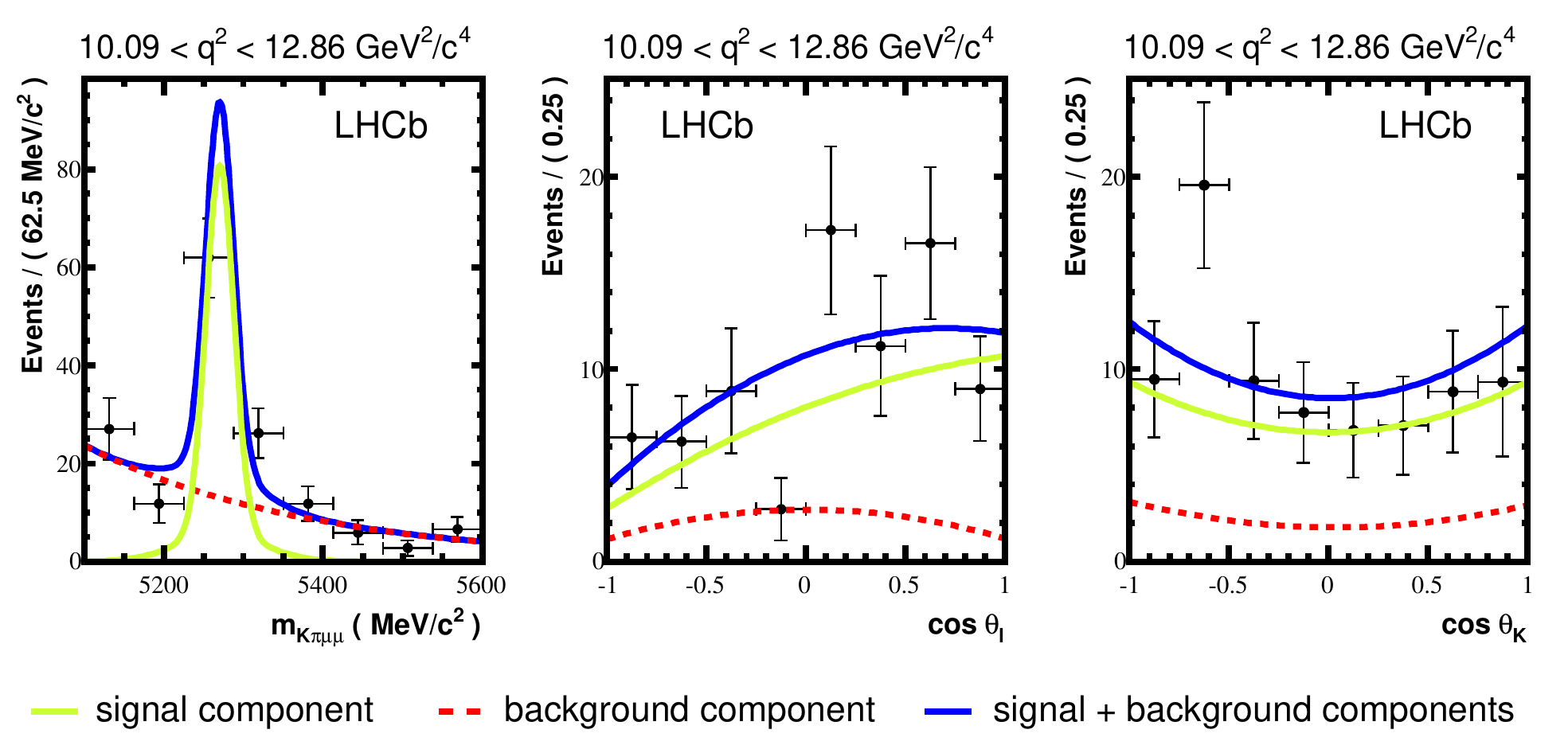}}
{\includegraphics[width=0.9\textwidth]{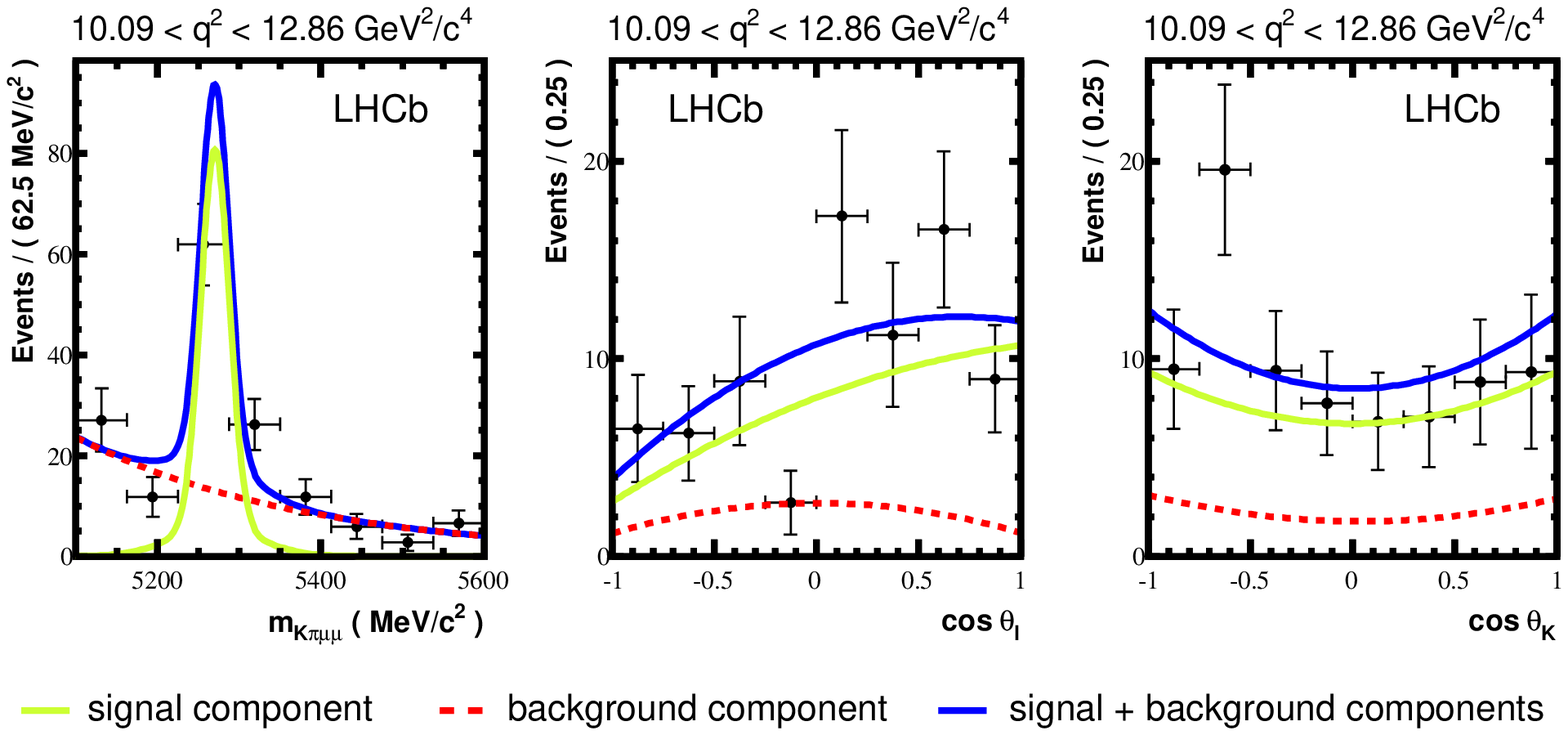}}
\ifthenelse{\boolean{pdflatex}}
{\includegraphics[width=0.9\textwidth]{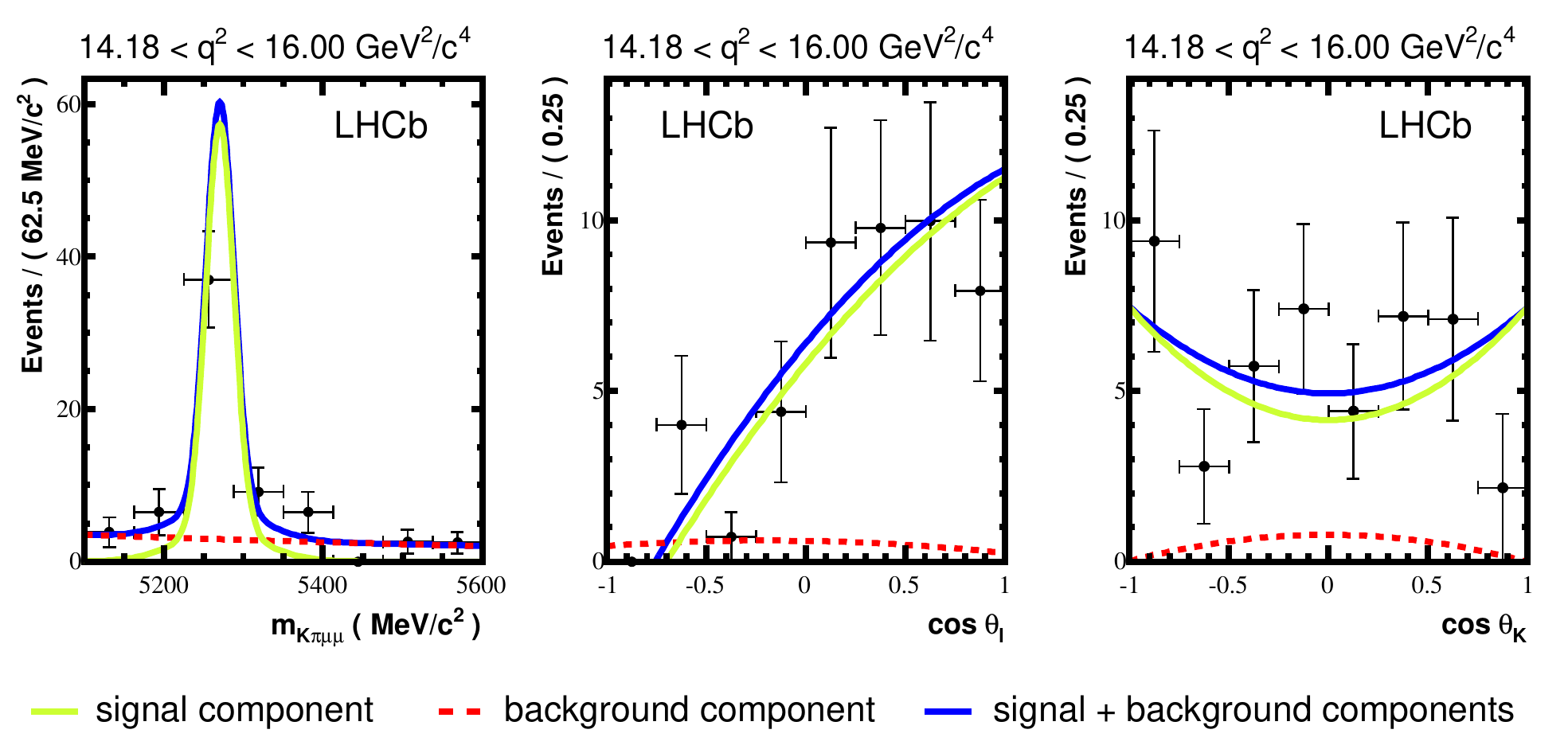}}
{\includegraphics[width=0.9\textwidth]{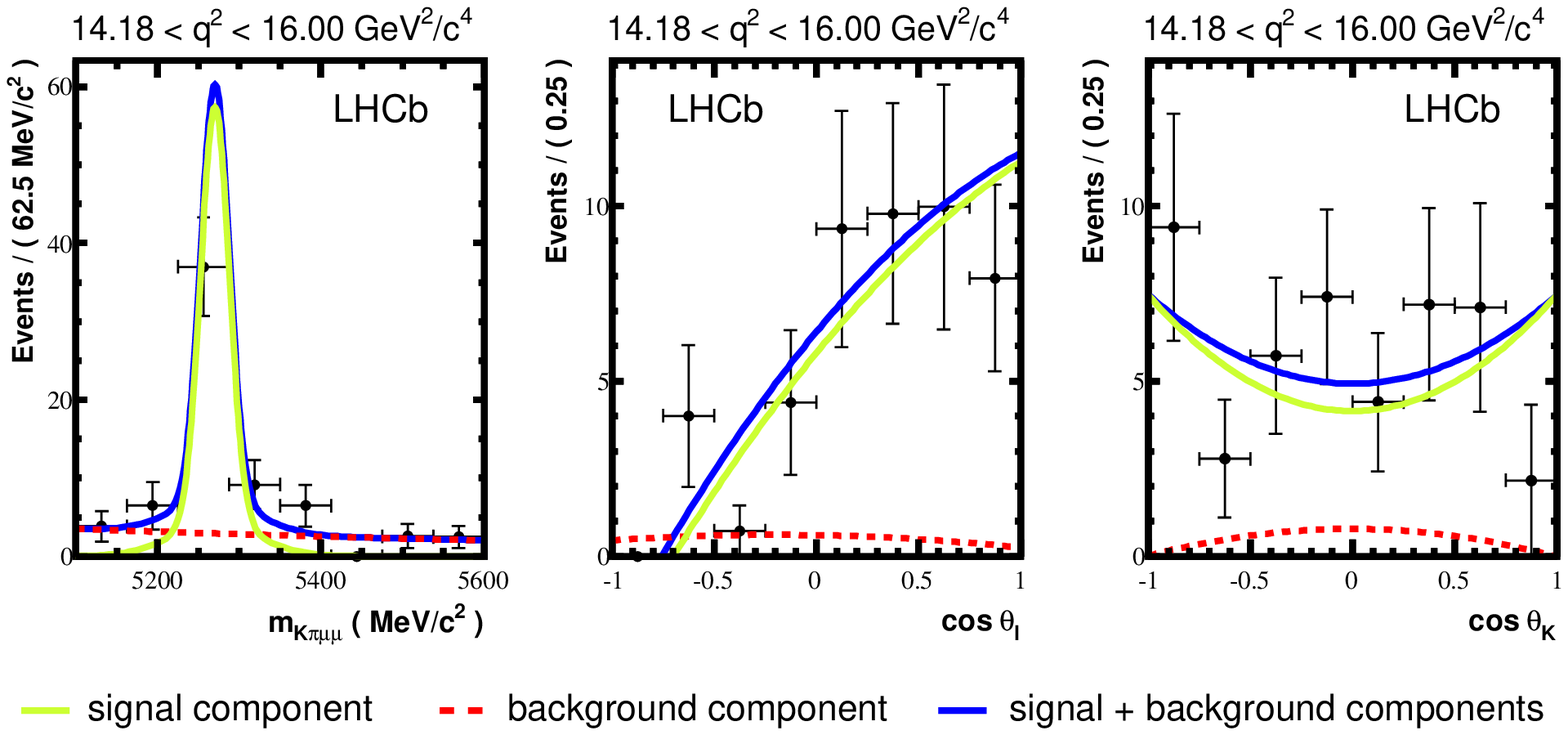}}
\ifthenelse{\boolean{pdflatex}}
{\includegraphics[width=0.9\textwidth]{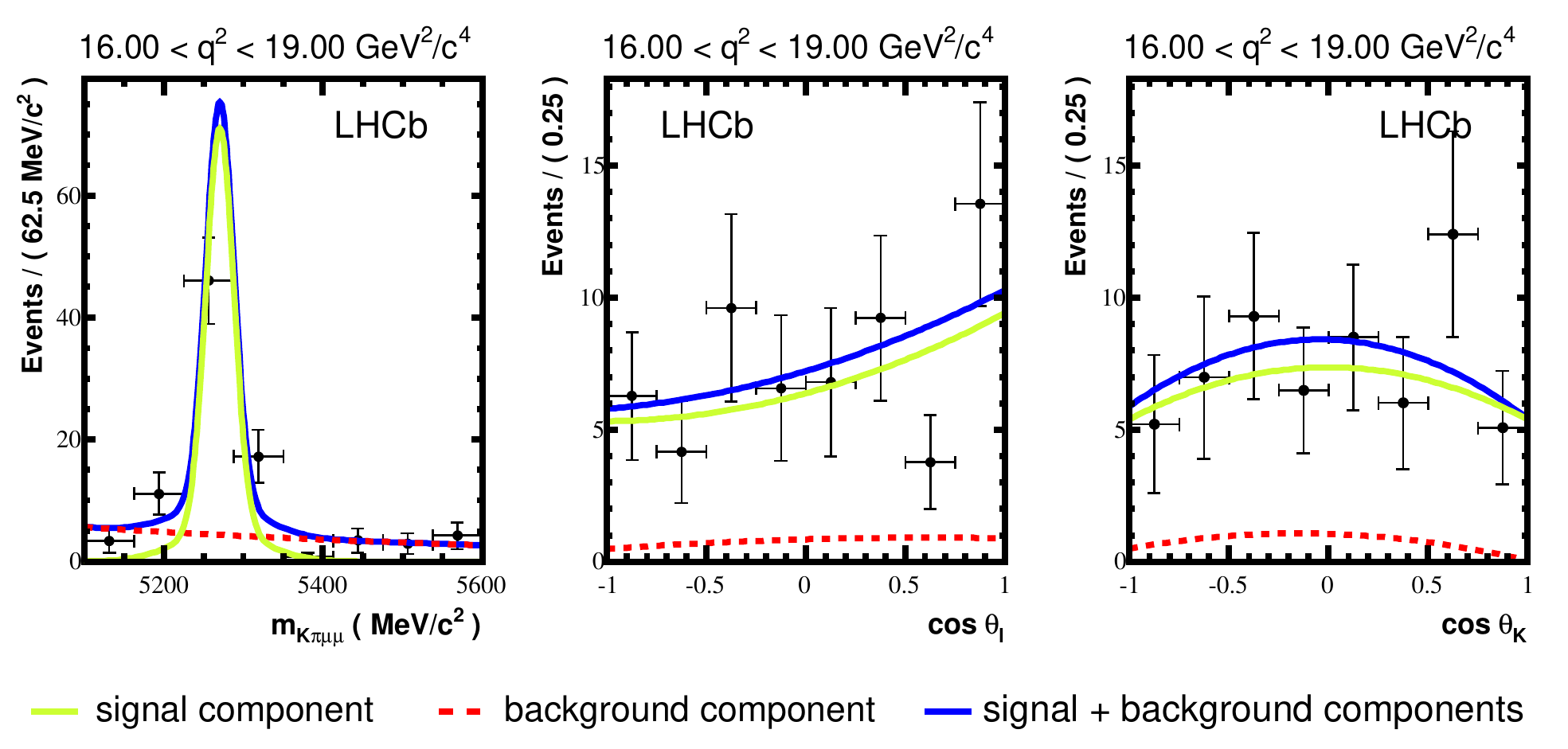}}
{\includegraphics[width=0.9\textwidth]{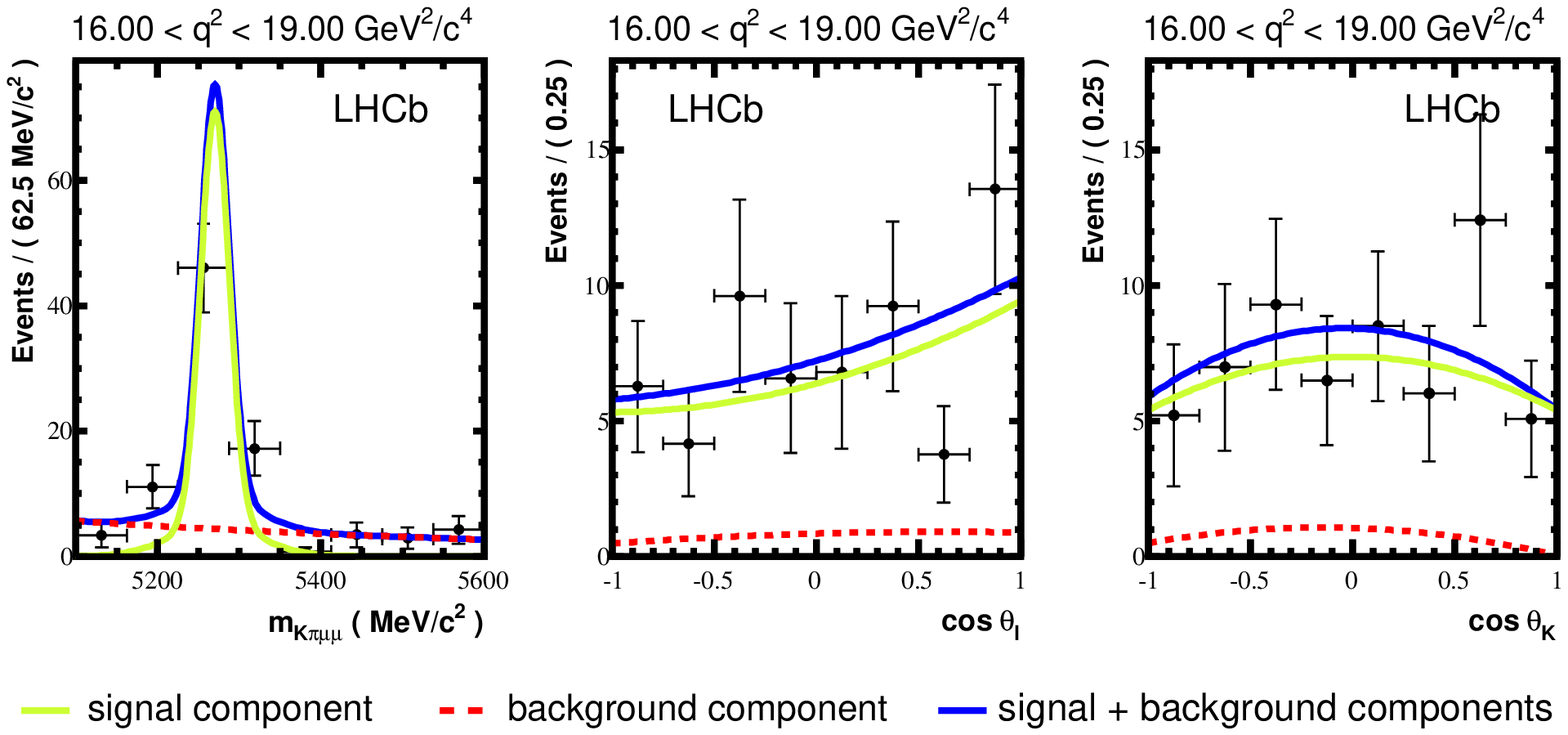}}
\caption{Fit projections for $m_{K\pi\mu\mu}$, \ctl and \ctk for the \qsq bins: $10.09 < q^{2} < 12.86$, $14.18 < q^{2} < 16.00$ and $16.00 < q^{2} < 19.00\gev^{2}/c^4$.}
\end{figure*}

\end{document}